# Artificial Intelligence Forecasting of Covid-19 in China

Zixin Hu[1,2], Qiyang Ge[3], Shudi Li[4], Li Jin[1,2] and Momiao Xiong[4,*]

[1] State Key Laboratory of Genetic Engineering and Innovation Center of Genetics and Development, School of Life Sciences, Fudan University, Shanghai, China.

[2] Human Phenome Institute, Fudan University, Shanghai, China.

[3] The School of Mathematic Sciences, Fudan University, Shanghai, China.

[4] Department of Biostatistics and Data Science, School of Public Health, The University of Texas Health Science Center at Houston, Houston, TX 77030, USA.

**Running Title**: Real Time Forecasting of Covid-19 in China

[*]Address for correspondence and reprints: Dr. Momiao Xiong, Department of Biostatistics and Data Science, School of Public Health, The University of Texas Health Science Center at Houston, P.O. Box 20186, Houston, Texas 77225, (Phone): 713-500-9894, (Fax): 713-500-0900, E-mail: Momiao.Xiong@uth.tmc.edu.

# Abstract

## BACKGROUND

An alternative to epidemiological models for transmission dynamics of Covid-19 in China, we propose the artificial intelligence (AI)-inspired methods for real-time forecasting of Covid-19 to estimate the size, lengths and ending time of Covid-19 across China.

## METHODS

We developed a modified stacked auto-encoder for modeling the transmission dynamics of the epidemics. We applied this model to real-time forecasting the confirmed cases of Covid-19 across China. The data were collected from January 11 to February 27, 2020 by WHO. We used the latent variables in the auto-encoder and clustering algorithms to group the provinces/cities for investigating the transmission structure.

## RESULTS

We forecasted curves of cumulative confirmed cases of Covid-19 across China from Jan 20, 2020 to April 20, 2020. Using the multiple-step forecasting, the estimated average errors of 6-step, 7-step, 8-step, 9-step and 10-step forecasting were 1.64%, 2.27%, 2.14%, 2.08%, 0.73%, respectively. We predicted that the time points of the provinces/cities entering the plateau of the forecasted transmission dynamic curves varied, ranging from Jan 21 to April 19, 2020. The 34 provinces/cities were grouped into 9 clusters.

## CONCLUSIONS

The accuracy of the AI-based methods for forecasting the trajectory of Covid-19 was high. We predicted that the epidemics of Covid-19 will be over by the middle of April. If the data are reliable and there are no second transmissions, we can accurately forecast the transmission dynamics of the Covid-19 across the provinces/cities in China. The AI-inspired methods are a powerful tool for helping public health planning and policymaking.

## Introduction

In the beginning of December, 2019, Covid-19 virus that slipped from animals to humans in Wuhan city, China caused an outbreak of respiratory illness. A number of the statistical, dynamic and mathematical models of the Covid-19 outbreak including the SEIR model have been developed to analyze its transmission dynamics[1,2,3,4,5]. Although these epidemiological models are useful for estimating the dynamics of transmission, targeting resources and evaluating the impact of intervention strategies, the models require parameters and depend on many assumptions. Unlike system identification in engineering where the parameters in the models are estimated using real data, at the outbreak, estimated parameters using real time data are not readily available[6,7] Most analyses used hypothesized parameters and hence do not fit the data very well. The accuracy of forecasting the future cases of Covid-19 using these models may not be very high. Timely interventions are needed to control the serious impacts of Covid-19 on health.

To overcome limitations of the epidemiological model approach, and assist public health planning and policy making, we develop an AI based method for real time forecasting of the new and cumulative confirmed cases of Covid-19 in total and provinces/ cities across China. We also forecast the possible trend and plateau of Covid-19 transmission in China and group the provinces/cities into clusters according to the dynamic patterns of Covid-19 transmission. The analysis is based on the surveillance data of the confirmed Covid-19 cases in China up to February 18, 2020.

## Methods

**Data Sources**

Data on the confirmed cases of Covid-19 from January 11, 2020 to January 20, 2020, and from January 21, 2020 to February 27, 2020, were from Surging News Network (https://www.thepaper.cn/) and WHO (https://www.who.int/emergencies/diseases/novel-coronavirus-2019/situation-reports), respectively. WHO took the lab confirmed case as the confirmed cases from January 21, 2020 to February 13, 2020 and took the sum of the number of the clinical confirmed and lab conformed cases as the number of the confirmed cases after February 14, 2020. The numbers of the clinical confirmed and the lab confirmed cases in Hubei Province, China on February 14, was 15,384 and 36,602, respectively. Therefore, the number of the confirmed cases in Hubei Province before February 14, 2020 was adjusted by the formula:

the number of the confirmed cases = the number of the lab confirmed cases $\times \frac{(15384+36602)}{366602}$.

Data included the total numbers of the accumulated and new confirmed cases in all of China and the numbers of the accumulated and new confirmed cases across 31 Provinces/Cities in mainland China and three other regions (Hong Kong, Macau and Taiwan) in China. The data were organized in a matrix with the rows representing the whole China and province/city and columns representing the number of the new confirmed cases of each day.

The confirmed cases of each province/city were a time series. Let $t_{ij}$ be the number of the confirmed cases of the $j^{th}$ day within the $i^{th}$ province/city. Let $Z$ be a $34 \times m$ dimensional matrix. The element $Z_{ij}$ is the number of the confirmed new cases of Covid-19 on the $j^{th}$ day, starting with January 11, 2020 in the $i^{th}$ city.

**Modified Auto-encoder for Modeling Time Series**

Modified auto-encoders (MAE)[8,9] were used to forecast the number of the accumulative and new confirmed cases of Covid-19. Unlike the classical auto-encoder where the number of nodes in the layers usually decreases from the input layer to the latent layers, the numbers of the nodes in the input, the first latent layer, the second latent layer and output layers in the MAE were 8, 32, 4 and 1, respectively (Figure 1). We view a segment of time series with 8 days as a sample of data and take 128 segments of time series as the training samples. One element from the dada matrix $Z$ is randomly selected as a start day of the segment and select its 7 successive days as the other days to form a segment of time series. Let $i$ be the index of the segment and $j_i$ be the column index of the matrix $Z$ that was selected as the starting day. The $i^{th}$ segment time series can be represented as $\{Z_{j_i}, Z_{j_i+1}, \ldots, Z_{j_i+7}\}$. Data were normalized to $X_{j_i+k} = \frac{Z_{j_i+k}}{S}, k = 0, 1, \ldots, 7$, where $S = \frac{1}{8}\sum_{k=0}^{7} Z_{j_i+k}$. Let $Y_i = \frac{Z_{j_i+8}}{S}$ be the normalized number of cases to forecast. If $S = 0$, then set $Y_i = 0$. The loss function was defined as

$$L = \sum_{i=1}^{128} W_i (Y_i - \hat{Y}_i)^2,$$

where $Y_i$ was the observed number of the cases in the forecasting day of the $i^{th}$ segment time series and $\hat{Y}_i$ was its forecasted number of cases by the MAE, and $W_i$ were weights. If $j_i$ was in the interval [1, 12], then $W_i = 1$. If $j_i$ was in the interval [13, 24], then $W_i = 2$, etc. The back propagation algorithm was used to estimate the weights and bias in the MAE. Repeat training processes 5 times. The average forecasting $\hat{Y}_i, i = 1, \ldots, 34$ will be taken as a final forecasted number of the accumulated confirmed cases for each province/city.

**Forecasting Procedures**

The trained MAE was used for forecasting the future number of the confirmed cases of Covid-19 for each province/city. Consider the $i^{th}$ province/city. Assume that the number of new confirmed cases of Covid-19 on the $j^{th}$ day that needs to be forecasted is $x_{ij}$. Let $H$ be a $34 \times 8$ dimensional matrix and $h_{il} = x_{ij-9+l}, i = 1, \ldots, 34,$ and $l = 1, \ldots, 8$. Let $g_i = \frac{1}{8}\sum_{l=1}^{8} h_{il}, i = 1, \ldots, 34$ be the average of the $i^{th}$ row of the matrix $H$. Let $U$ be the normalized matrix of $H$ where $u_{il} = \frac{h_{il}}{g_i}, i = 1, \ldots, 34,$ and $l = 1, \ldots, 8$. The output of the MAE is the forecasted number of the new confirmed cases and is denoted as $\hat{v}_i = f(u_{i1}, \ldots, u_{i8}, \theta), i = 1, \ldots, 34$, where $\theta$ represented the estimated parameters in the trained MAE. The one-step forecasting of the number of the new confirmed cases of Covid-19 for each city is given by $\hat{Y}_i = \hat{v}_i g_i, i = 1, \ldots, 34$.

The recursive multiple-step forecasting involved using a one-step model multiple times where the prediction for the preceding time step was used as an input for making a prediction on the following time step. For example, for forecasting the number of the new confirmed cases for the one more next day, the predicted number of new cases in one-step forecasting would be used as an observational input in order to predict day 2. Repeat the above process to obtain the two-step forecasting. The summation of the final forecasted number of the new confirmed cases for each province/city was taken as the prediction of the total number of the new confirmed cases of Covid-19 in China.

**Clustering**

The values of the latent variables in the second latent layer of the MAE for each province/city were extracted. For each province/city, a $34 \times 4$ dimensional latent matrix $A$ were formed. The largest single value $\lambda$ of the latent matrix $A$ was obtained via single value decomposition. We performed five-time trainings and obtained five largest single values. For each province/city, we formed a feature vector that consisted of the five largest single values $\lambda s$, the starting day and the forecasted end day of the Covid-19 outbreak, the day, the number of new confirmed cases reaching the maximum, the largest number of the forecasted new confirmed cases and the number of the forecasted accumulated confirmed cases of Covid-19 in the respective province/city. The k-means algorithms were performed on the 34 feature vectors to group provinces/cities into clusters.

**Results**

Figure 2 plotted the total number curves of the reported and forecasted cumulative and new confirmed cases of Covid-19 in China as a function of days. The reported cases were from January 11, 2020 to February 27, 2020. A total number of 47 days' data were available. We began to forecast on February 20, 2020. Figure 2 showed that the forecasting curve was close to the reported curve. From Figure 2 it can be observed that the curve of the new confirmed cases reached the 5,236 peak on February 5, 2020 and decreased to zero on April 20 (forecasting). The potential cumulative confirmed cases of Covid-19 in China was observed to reach the plateau (83,401) on April 20, 2020. Figure S1 plotted the national reported and fitted curves of the cumulative confirmed cases in China from January 20, 2020 to February 27, 2020. To further examine the accuracy of forecasting, Table 1 where the data from January 20, 2020 to February 27, 2020 were used to fit the MAE model. In the table, we listed 1 day-step to 10 day-step forecasting errors, respectively, i.e., the errors of using the current reported cases to forecast future s day cases. Table 1 showed that the average errors of the forecasting did not strictly increase as the number-steps for forecasting increased due to fluctuations of the data. Forecasting accuracy was very high.

Figure 3 presented the forecasted curves of cumulative confirmed cases of Covid-19 across 31 province/cities in mainland China and three other regions (Hong Kong, Macau and Taiwan) in China as a function of days from January 11, 2020 to April 20, 2020. We observed from Figure 3 that the times of different provinces/cities entering the plateau will be different, ranging from Jan 21, 2020 to April 20, 2020. Xizang first entered the plateau at Jan 21 followed by Macao on Jan 26 and Qinghai on Jan 28. Hubei province will be one of the last to enter the plateau (70,019) around April 20, 2020, most provinces/cities will enter the plateau around middle of March. We can also observe the different shapes of the curves among the provinces/cities, which imply different dynamic patterns of the transmission of Covid-19 across the provinces/cities.

Many factors such as size of outflow population from Wuhan to each affected provinces/cities, interventions, geographic locations, economic and social activities, environmental heterogeneity and healthcare facility affect disease transmission dynamics across the country. Clustering its temporal dynamics will provide numerous insights on patterns of propagation of Covid-19. To further capture the dynamic pattern of Covid-19 spread across the provinces/cities, we presented Figure 4 where 34 provinces/cities formed nine clusters. Since the spread of Covid-19 was influenced by the pattern of contacts between individuals, clusters partially showed geographic structure.

For example, Hubei province, a major source of Covid-19, formed a cluster. Followed by the Anhui, Guangdong, Henan, Hunan, Jiangsu, Jiangxi and Zhejiang, which are in areas surrounding pathogen sources (Hubei province). The Inner Mongolia, Xizang and Qinghai cluster had the lowest number of cases and are in areas far away from the sources of Covid-19. In addition to the geographic locations that affected the transmission of Covid-19, the healthcare resources and economic and social activities may also affect the transmission dynamics of Covid-19 as the cluster of Shanghai, Hongkong, Beijing, Shandong and Sichuan. And Fujian Guangxi, Hebei, Jilin and Tianjin formed a cluster, but these five provinces/cities are not located in the same geographic area. They may have similar economic relationships with Wuhan, healthcare resources and take similar interventions to control of the spread of Covid-19.

**Discussion**

As an alternative to epidemiologic transmission model, we used MAE to forecast the real-time trajectory of the transmission dynamics and generate the real-time forecasts of Covid-19 across the provinces/cities in China. The results showed that the accuracies of prediction and subsequently multiple-step forecasting were high. Our experience revealed that forecasting improves when the training time was longer. We estimated the potential time points of decreasing growth of new confirmed case curves across 34 provinces/cities, the lengths of the Cov-19 epidemics across China, and the times when the number of accumulated confirmed cases of Covid-19 would reach the plateau of their accumulate case curves. If the data are reliable and there will be no second transmission, the MAE models predicted that the Covid-19 outbreak in China might be over in the middle of April. In this study, we used cluster analysis to group 34 provinces/cities into 9 clusters and explore their geographic and healthcare resource structures.

The MAE models allow inputting the interventions information and investigating the impact of interventions on the size of the virus outbreak and end time of the virus outbreak. However, we have not explored such functions of the MAE due to lack of data. Similar to epidemiologic transmission dynamic models, the MAE can also be used for simulations. The trained MAE can well approximate many dynamic processes. Using the hypothesized initial sizes of the epidemic outbreak, we can use the MAE with known parameters and architecture to estimate the sizes of outbreak in the future and simulate the impact of the interventions on the sizes and severity of the epidemics. Complimentary to a model approach to transmission dynamics of virus outbreaks, the data driven AI-based methods provide real time forecasting tools for tracking, estimating the trajectory of epidemics, assessing their severity, predicting the lengths of epidemics and assisting government and health workers to make plan and good decisions.

**Legend**

**Figure 1.** Architecture of a MAE.

**Figure 2.** The national reported and forecasted curves of the cumulative and new confirmed cases of Covid-19 in China as a function of days from January 11, 2020 to April 20, 2020.

**Figure 3.** The forecasted curves of the cumulative confirmed cases of Covid-19 across 34 province/cities in China as a function of days from January 11, 2020 to April 20, 2020.

**Figure 4.** The clusters that were grouped by features extracted from the MAE and the cumulative confirmed case time series of Covid-19 across 31 provinces/cities in mainland China and three other regions in China formed 9 clusters.

**Table 1.** Errors of forecasting the national cumulative confirmed cases in China.

**Supplementary Figure Legend**

**Figure S1.** The national reported and fitted curves of the cumulative confirmed cases of Covid-19 in China from January 11, 2020 to February 27, 2020, where the red curve was the reported and green curve was the fitted.

**Figure 1.** Architecture of a MAE

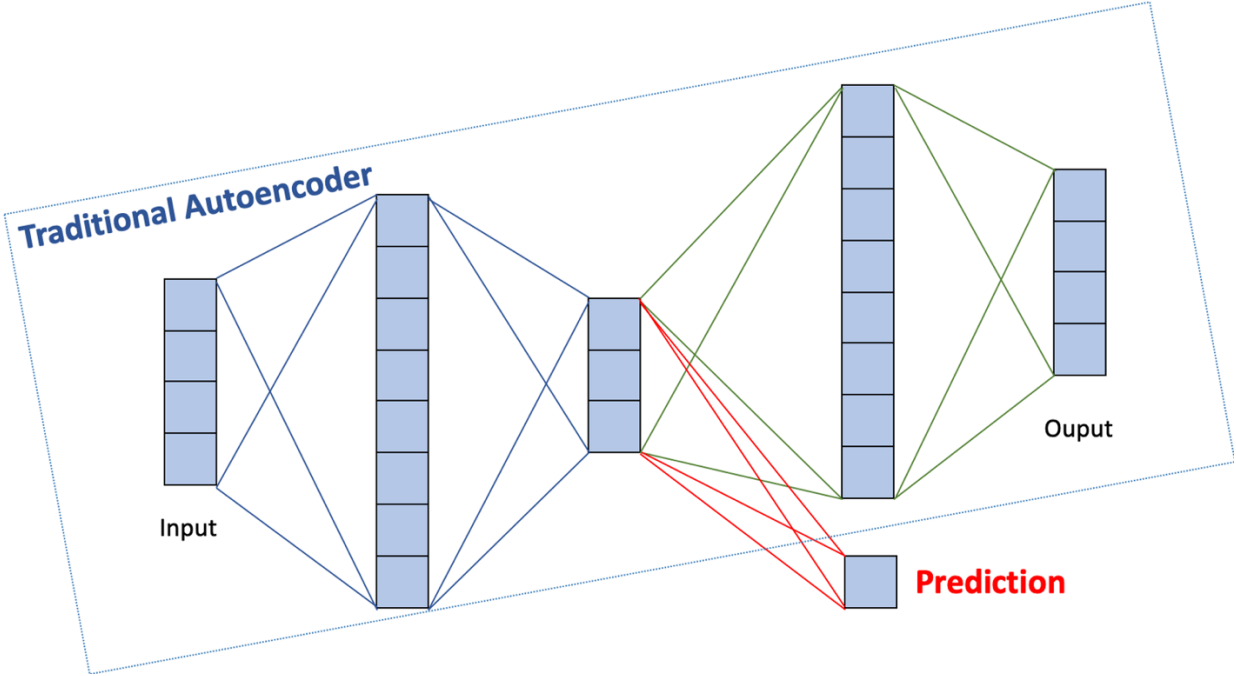

**Figure 2.** The national reported and forecasted curves of the cumulative and new confirmed cases of Covid-19 in China as a function of days from January 11, 2020 to April 20, 2020.

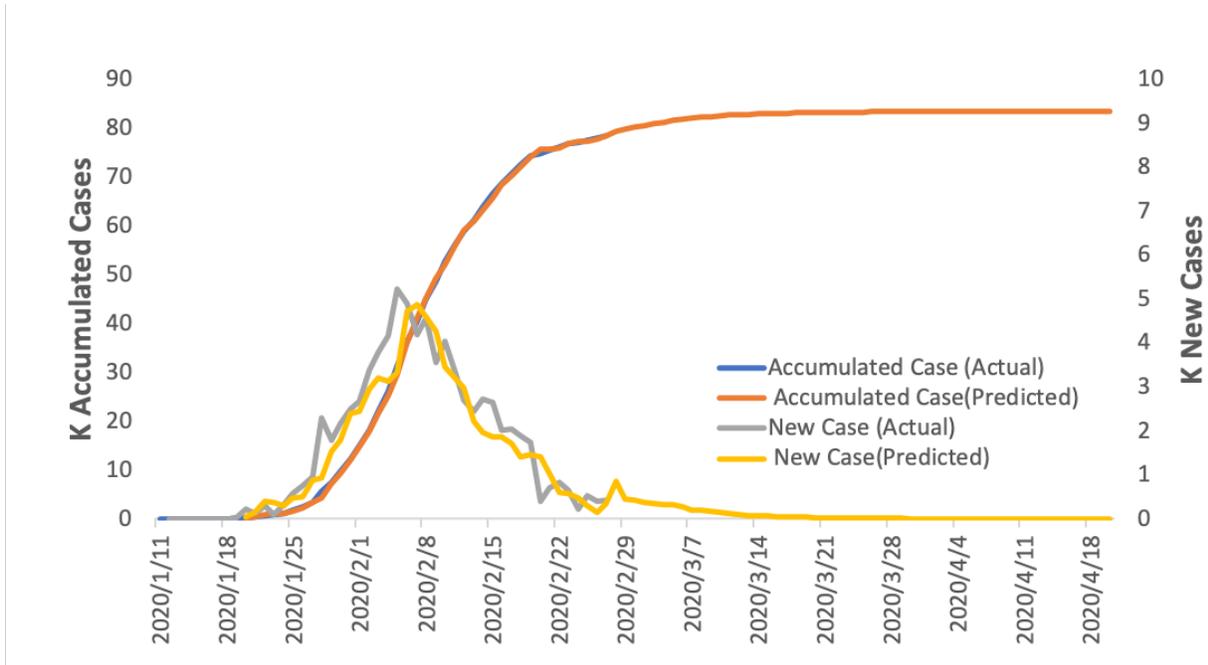

**Figure 3.** The forecasted curves of the cumulative confirmed cases of Covid-19 across 34 province/cities in China as a function of days from January 11, 2020 to April 20, 2020.

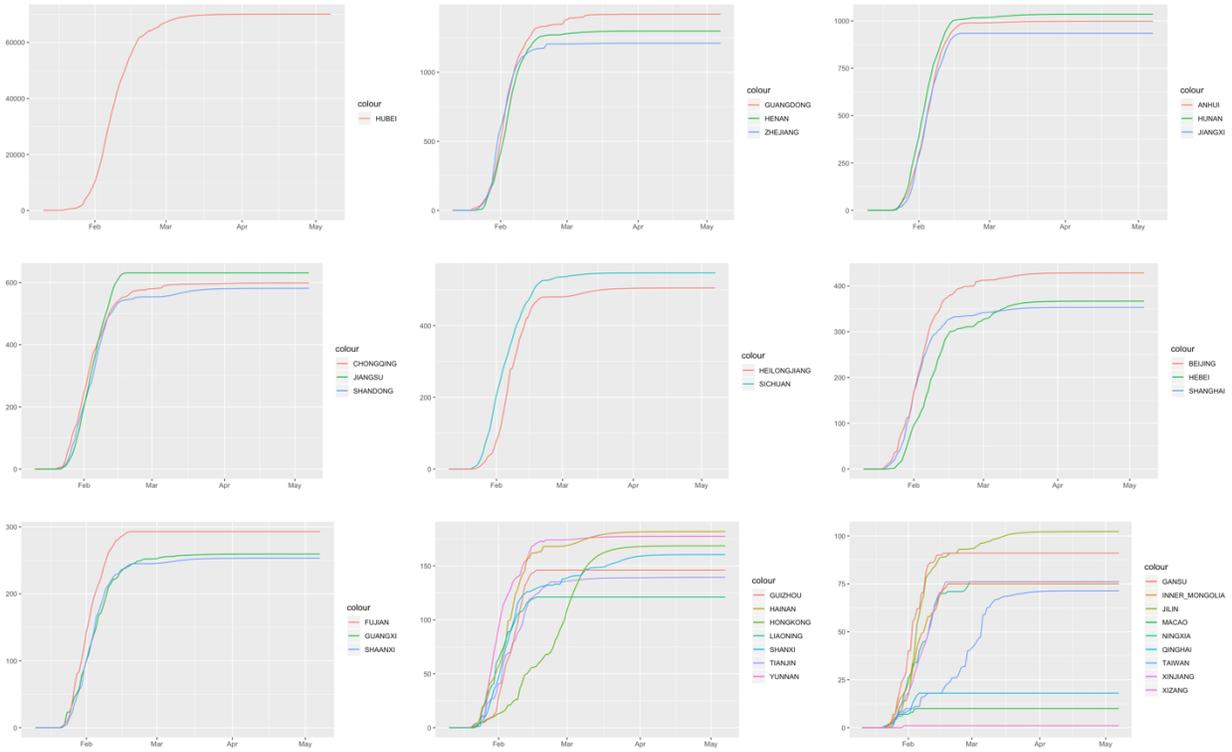

**Figure 4.** The clusters that were grouped by features extracted from the MAE and the cumulative confirmed case time series of Covid-19 across 31 provinces/cities in mainland China and three other regions in China formed 9 clusters.

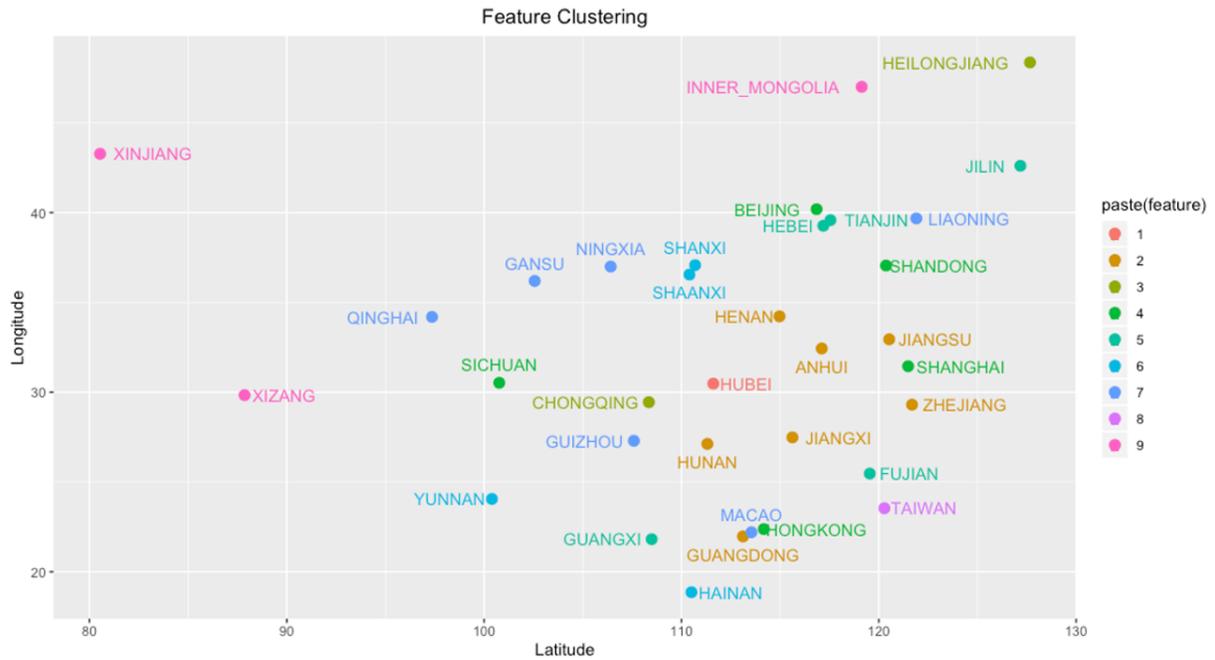

Cluster1: HUBEI
Cluster2: ANHUI, GUANGDONG, HENAN, HUNAN, JIANGSU, JIANGXI, ZHEJIANG
Cluster3: CHONGQING, HEILONGJIANG
Cluster4: BEIJING, HONGKONG, SHANDONG, SHANGHAI, SICHUAN
Cluster5: FUJIAN, GUANGXI, HEBEI, JILIN, TIANJIN
Cluster6: HAINAN, SHAANXI, SHANXI, YUNNAN
Cluster7: GANSU, GUIZHOU, LIAONING, MACAO, NINGXIA, QINGHAI
Cluster8: TAIWAN
Cluster9: INNER_MONGOLIA, XINJIANG, XIZANG

**Table 1.** Errors of forecasting the national cumulative confirmed cases in China.

| Date | Actual | 1-step prediction | 1-step error | 2-step error | 3-step error | 4-step error | 5-step error | 6-step error | 7-step error | 8-step error | 9-step error | 10-step error |
|---|---|---|---|---|---|---|---|---|---|---|---|---|
| 2020/2/18 | 72,528 | 71,757 | -1.06% | | | | | | | | | |
| 2020/2/19 | 74,280 | 74,005 | -0.37% | -1.34% | | | | | | | | |
| 2020/2/20 | 74,675 | 75,564 | 1.19% | 1.31% | -0.06% | | | | | | | |
| 2020/2/21 | 75,569 | 76,685 | 1.48% | 1.36% | 1.49% | 0.12% | | | | | | |
| 2020/2/22 | 76,392 | 76,305 | -0.11% | 1.49% | 1.41% | 1.81% | 0.15% | | | | | |
| 2020/2/23 | 77,042 | 77,827 | 1.02% | 0.02% | 1.58% | 1.39% | 2.30% | 0.13% | | | | |
| 2020/2/24 | 77,262 | 79,837 | 3.33% | 2.18% | 0.83% | 2.53% | 2.46% | 3.30% | 1.08% | | | |
| 2020/2/25 | 77,780 | 79,155 | 1.77% | 3.01% | 2.14% | 0.47% | 2.28% | 2.36% | 3.15% | 0.76% | | |
| 2020/2/26 | 78,191 | 77,957 | -0.30% | 1.61% | 2.70% | 2.15% | 0.28% | 2.20% | 2.49% | 3.21% | 0.80% | |
| 2020/2/27 | 78,630 | 78,646 | 0.02% | -0.53% | 1.49% | 2.47% | 2.22% | 0.21% | 2.37% | 2.46% | 3.37% | 0.73% |
| **Average Absolute Error** | | | 1.07% | 1.43% | 1.46% | 1.56% | 1.62% | 1.64% | 2.27% | 2.14% | 2.08% | 0.73% |

**Figure S1.** The national reported and fitted curves of the cumulative confirmed cases of Covid-19 in China from January 11, 2020 to February 27, 2020, where the red curve was the reported and green curve was the fitted.

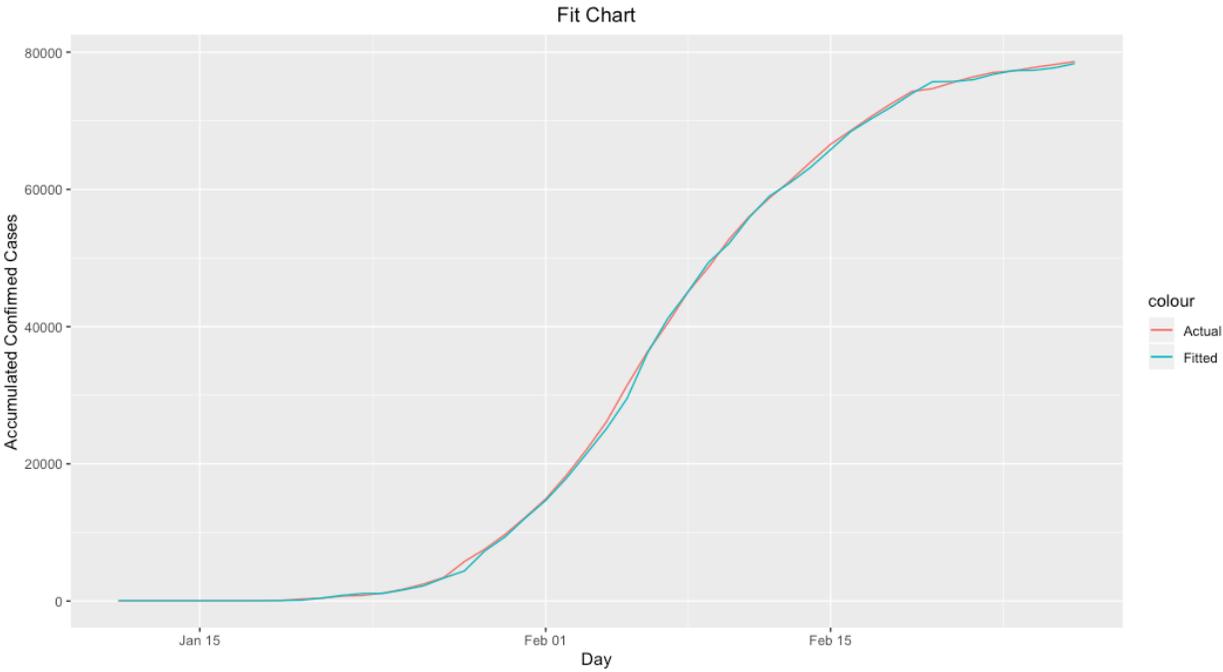